\def\be{\begin{equation}}
\def\ee{\end{equation}}
\def\ba{\begin{array}}
\def\p{\prime}
\def\ea{\end{array}}

\def\Rb{{I\!\! R}}

\def\Cb{\ \hbox{\vrule width 0.6pt height 6pt depth 0pt
		      \hskip -3.5 pt} C}
\def\cb{\ \hbox{\vrule width 0.6pt height 4pt depth 0pt
		      \hskip -3.5 pt} C}

\documentstyle[12pt]{article}
\begin{document}
\parskip=4pt
\parindent=18pt
\baselineskip=22pt
\setcounter{page}{1}
\centerline{\Large\bf Many Body Problems with ``Spin"-Related}
\centerline{\Large\bf Contact Interactions}
\vspace{5ex}

\centerline{S. Albeverio$^{1,2}$, S-M. Fei$^{1,3}$ and P. Kurasov$^4$}

\parskip=0pt
\parindent=30pt
\baselineskip=16pt
\vskip 1 true cm

$^1$ Institut f\"ur Angewandte Mathematik, Universit\"at Bonn, D-53115 Bonn

~~Fakult\"at f\"ur Mathematik, Ruhr-Universit\"at Bochum, D-44780 Bochum

$^2$ SFB 256; SFB 237; BiBoS; CERFIM (Locarno); Acc.Arch., USI (Mendrisio)

$^3$ Institute of Physics, Chinese Academy of Science, Beijing

$ ^4 $ Dept. of Math., Stockholm University, 10691 Stockholm, Sweden

~~Dept. of Math., Lule{\aa} University, S-97187 Lule{\aa}, Sweden

~~St. Petersburg University, 198904 St. Petersburg, Russia

\parindent=18pt
\parskip=6pt

\begin{center}
\begin{minipage}{5in}
\vspace{3ex}
\centerline{\large Abstract}
\vspace{4ex}
We study quantum mechanical systems with ``spin"-related
contact interactions in one dimension. The boundary conditions
describing the contact interactions are dependent on the spin states of
the particles. In particular we investigate the integrability of
$N$-body systems with $\delta$-interactions and point spin couplings.
Bethe ansatz solutions, bound states and scattering matrices are
explicitly given. The cases of generalized separated boundary condition
and some Hamiltonian operators corresponding to special spin related
boundary conditions are also discussed.
\end{minipage}
\end{center}

\newpage
\baselineskip=22pt

Quantum mechanical solvable models describing a particle moving in a
local singular potential concentrated at one or a discrete number of points
have been extensively discussed in the literature, see e.g.
\cite{agh-kh,AKbook,gaudin} and references therein.
One dimensional problems with contact interactions at, say, the origin
($x=0$) can be  characterized by
the boundary conditions imposed on the (scalar) wave function $\varphi$ at
$x=0$. The history of this problem is well described
in \cite{agh-kh,AKbook}.
It was suggested to divide these conditions into two disjoint
families:
separated and nonseparated boundary conditions, corresponding
to the cases where the perturbed operator is equal to
the orthogonal sum of two self--adjoint operators in $ L_{2}
(-\infty,0] $ and $L_{2} [0,\infty) $ and when this representation
is impossible, respectively.
Classification of one dimensional point interactions
in terms of singular perturbations is given in \cite{kurasov}.
In the present paper we are interested in model few-body problems with
pairwise interactions given by such potentials.
The first model of this type
with the pairwise interactions determined by delta
functions was suggested and investigated
 by J.B. McGuire and C.A. Hurst
  \cite{mcguire,mcguire2,mcguire3,mcguire4,mcguire5}.
 The eigenfunctions for the
system of identical particles interacting via delta potentials
are given by Bethe Ansatz.
Intensive studies of this model applied to statistical
mechanics (particles having boson or fermion statistics)
 by C.N.Yang and his collaborators
lead to the famous Yang-Baxter equation \cite{y1,y,y2}.
It has been shown in \cite{buslaev,revmath,makarov} that
$N$-particle systems with three-body interactions do not
have eigenfunctions given by Bethe Ansatz.
In \cite{ADF} the integrability of one dimensional systems of $N$
identical particles with general contact interactions described by
the boundary conditions imposed on the wave function was
investigated.
 It was shown that the $N$-particle system
satisfies a Yang-Baxter relation not only
in the $\delta$-interaction case, but also for two other
one parameter (sub)families, one with nonseparated boundary conditions
and another with separated boundary conditions.
This fact is not surprising, since the Yang-Baxter equation
has been derived for particles with boson or fermion statistics.
Suppose that the system of $ N $ particles satisfies one of
these statistics. Then the eigenfunction equation can be
reduced to an equation in the sector
$ x_{1} \leq x_{2} \leq \ldots \leq x_{N}, $
since the value of the total wave function in the whole
space $ {\bf R}^{N} $
can be reconstructed using symmetry properties of this
wave function.
The boundary conditions on the total wave function
are transferred into certain conditions
at the boundaries of the sector for the
reduced wave function.
In fact all three families of boundary conditions obtained
in \cite{ADF} correspond to one reduced problem.
Hence as far as particles with statistics are concerned
the only difference between the three families is
due to the symmetry properties of the wave function, i.e.
the rule how the total wave function can be reconstructed
from the reduced one.
Considering particles without any statistics the eigenfunctions
corresponding to the boundary conditions from the three
one parameter families the eigenfunctions can be calculated
using Bethe Ansatz.
In \cite{AFK} it is shown
that not only the models satisfying the Yang-Baxter equation have
eigenfunctions of the type of those constructed following Bethe Ansatz.
This is possible, since to derive Yang-Baxter equation from
Bethe Ansatz one has to use symmetry properties of the wave
function determined by the statistics.
The family of such model operators is described by
two real parameters. One of these parameters
is redundant in the sense that the operators corresponding
to different values of this parameter are unitary equivalent.
It is shown in \cite{AFK}
that the redundant parameter can be interpreted as
the amplitude of a singular gauge field.
Note that this parameter can play an important role
for nonstationary problems.
A similar problem has been studied in \cite{coutinho}
but it was wrongly concluded there that the family of
such models having eigenfunctions given by Bethe Ansatz
coincides with the family of models satisfying the Yang-Baxter
equation. This point has been already clarified in
\cite{AFK}.

The family of point interactions for the one dimensional
Schr\"odinger operator $ - \frac{d^2}{dx^2} $
can be described by unitary $ 2 \times  2 $ matrices
via von Neumann formulas for self--adjoint extensions
of symmetric operators, since the second derivative
operator restricted to the domain $ C_{0}^\infty ({\bf R}
\setminus \{ 0 \} ) $ has deficiency indices $ (2,2)$.
In what follows we are going to
consider only the self--adjoint nonseparated
 extensions that cannot be
presented as an orthogonal sum of two self--adjoint
operators acting in $ L_{2} (-\infty, 0] $ and
$ L_{2} [0, \infty) $.
The boundary conditions describing the self-adjoint extensions
have the following form
\begin{equation} \label{bound}
\left( \begin{array}{c}
\varphi\\
\varphi '\end{array} \right)_{0^+}
= e^{i\theta} \left(
\begin{array}{cc}
a & b \\
c & d \end{array} \right)
\left( \begin{array}{c}
\varphi\\
\varphi '\end{array} \right)_{0^-},
\end{equation}
where
\be\label{abcd}
ad-bc = 1,~~~~\theta, a,b,c,d \in \Rb.
\ee
$\varphi(x)$ is the scalar wave function of two particles with spin
$0$ and relative coordinate $x$. (\ref{bound}) also describes two particles
with spin $s$ but without any spin coupling between the particles when they
meet (i.e. for $x=0$), in this case $\varphi$ represents any one of the
components of the wave function. The values $\theta = b=0$, $a=d=1$ in
(\ref{bound})
correspond to the case of a positive (resp. negative) $\delta$-function
potential for $c>0$ (resp. $c<0$). For general $a,b,c$ and $d$, the
properties of the corresponding Hamiltonian systems have been studied in
detail, see e.g. \cite{abd,15b,adk,ch,kurasov,seba}.

For a particle with spin $s$, the wave function has $n=2s+1$ components.
Therefore two particles with contact interactions have a general boundary
condition described in the center of mass coordinate system by:
\be\label{BOUND}
\left( \begin{array}{c}
\psi\\
\psi '\end{array} \right)_{0^+}
=\left(
\begin{array}{cc}
A & B \\
C & D \end{array} \right)
\left( \begin{array}{c}
\psi\\
\psi '\end{array} \right)_{0^-},
\end{equation}
where $\psi$ and $\psi '$ are $n^2$-dimensional
column vectors, $A,B,C$ and $D$ are
$n^2\times n^2$ matrices. The boundary condition (\ref{BOUND})
can  include
not only the usual contact interaction between the particles,
but also a spin coupling
of the two particles if the matrices $ A, B, C, D $ are not diagonal.
These conditions are similar to those appeared
in \cite{albku,jabo} during the investigation of finite
rank singular perturbations of differential
operators.

The matrices $A,B,C$, and $D$ are subject to restrictions due to
the required symmetry condition of the Schr\"odinger operator. In
fact we should have, for any $u,v\in C^\infty(\Rb \setminus \{0\})$,
\be\label{symm}
\ba{l}
\displaystyle<-\frac{d^2}{dx^2}u,v>_{L_2(\Rb,\cb^n)}-
<u,-\frac{d^2}{dx^2}v>_{L_2(\Rb,\cb^n)}\\[4mm]
=<u '(0^+),v(0^+)>_{\cb^n}-<u(0^+),v '(0^+)>_{\cb^n}\\[4mm]
~~~-<u '(0^-),v(0^-)>_{\cb^n}+<u(0^-),v '(0^-)>_{\cb^n}=0.
\ea
\ee
>From (\ref{BOUND}) and (\ref{symm}) we get the following conditions:
\be\label{ABCD}
A^\dagger D-C^\dagger B=1,~~~B^\dagger D=D^\dagger B,~~~
A^\dagger C=C^\dagger A,
\ee
where $\dagger$ stands for the conjugate and transpose. Obviously
(\ref{bound}) is the special case of (\ref{BOUND}) at $s=0$.

In the following we study quantum systems with contact interactions
described by the boundary condition (\ref{BOUND}), in particular, $N$-body
systems with $\delta$-interactions.
We first consider two spin-$s$ particles with $\delta$-interactions.
The Hamiltonian is then of the form
\be\label{H}
H=(-\frac{\partial^2}{\partial x_1^2}-\frac{\partial^2}{\partial
x_2^2}){\bf I}_2
+2h\delta(x_1-x_2),
\ee
where ${\bf I}_2$ is the $n^2\times n^2$ identity matrix,
$h $
is an $n^2\times n^2$  Hermitian matrix.
If the matrix $h$ is proportional to the unit matrix
$ {\bf I}_{2}$, then $H$ is reduced to the usual
two-particle Hamiltonian with contact interactions but no spin
coupling.

Let $e_\alpha$, $\alpha=1,...,n$,
be the basis (column) vector with the $\alpha$-th component as $1$
and the rest components $0$. The wave function of the system (\ref{H})
is of the form
\be\label{psi}
\psi=\sum_{\alpha,\beta=1}^n\phi_{\alpha
\beta}(x_1,x_2)e_\alpha\otimes e_\beta.
\ee
In the center of mass coordinate system, $X=(x_1+x_2)/2$, $x=x_1-x_2$,
the operator (\ref{H}) has the form
\begin{equation}
H = - \left( \frac{1}{2} \frac{\partial^2}{\partial X^2}
+ 2 \frac{\partial ^2}{\partial x^2} \right)
{\bf I}_{2} + 2 h \delta (x).
\end{equation}
The functions $ \phi = \phi (x,X) $
 from the domain of this operator
satisfy
 the following boundary condition at $x=0$,
\be\label{b2}
\phi_{\alpha\beta} '(0^+,X)-\phi_{\alpha\beta} '(0^-,X)=
\sum_{\alpha,\beta=1}^n h_{\gamma\lambda,\alpha\beta}
\phi_{\gamma\lambda}(0,X),~~~
\phi_{\alpha\beta}(0^+,X)=\phi_{\alpha\beta}(0^-,X),~~~\alpha,\beta=1,...,n,
\ee
where the indices of the matrix $h$ are arranged as $11,12,...,1n;21,22,...,
2n;...;n1,n2,...,nn$. (\ref{b2}) is a special case of (\ref{BOUND}) for
$A=D={\bf I}_2$, $B=0$ and $C=h$. $h$ acts on the basis vector of particles $1$
and $2$ by $h e_\alpha\otimes e_\beta =\displaystyle
\sum_{\gamma,\lambda=1}^n h_{\alpha\beta,\gamma\lambda} e_\gamma\otimes
e_\lambda$.

According to the statistics
$\psi$ is symmetric (resp. antisymmetric) under the interchange of
the two particles if $s$ is an integer (resp. half integer).
Let $k_1$ and $k_2$ be the momenta of the two particles.
In the region $x_1<x_2$, in terms of Bethe hypothesis the
wave function has the following form
\be\label{w1}
\psi=u_{12}e^{i(k_1x_1+k_2x_2)}+u_{21}e^{i(k_2x_1+k_1x_2)},
\ee
where $u_{12}$ and $u_{21}$ are $n^2\times 1$ column matrices.
In the region $x_1>x_2$,
\be\label{w2}
\psi=(P^{12}u_{12})e^{i(k_1x_2+k_2x_1)}
+(P^{12}u_{21})e^{i(k_2x_2+k_1x_1)},
\ee
where according to the symmetry or antisymmetry conditions,
$P^{12}=p^{12}$ for bosons and $P^{12}=-p^{12}$ for fermions, $p^{12}$
being the operator on the $n^2\times 1$ column that interchanges
the spins of the two particles.
Substituting (\ref{w1}) and (\ref{w2}) into the boundary
conditions (\ref{b2}), we get
\be\label{a1}
\left\{
\begin{array}{l}
u_{12}+u_{21}
=P^{12}(u_{12}+u_{21}),\\
ik_{12}(u_{21}-u_{12})
=hP^{12}(u_{12}+u_{21})+ik_{12}P^{12}(u_{12}-u_{21}),
\end{array}\right.
\ee
where $k_{12}=(k_1-k_2)/2$. Eliminating the term $P^{12}u_{12}$ from
(\ref{a1}) we obtain the relation
\be\label{2112}
u_{21} = Y_{21}^{12} u_{12}~,
\ee
where
\be\label{a21a12}
Y_{21}^{12}=[2ik_{12} -h]^{-1}[2ik_{12}P^{12}+h].
\ee

For a system of $N$ identical particles with $\delta$-interactions,
the Hamiltonian is given by
\be\label{HN}
H=-\sum_{i=1}^N\frac{\partial^2}{\partial x_i^2}{\bf I}_N+\sum_{i<j}^N h_{ij}
\delta(x_i-x_j),
\ee
where ${\bf I}_N$ is the $n^N\times n^N$ identity matrix,
$h_{ij}$ is an operator acting on the $i$-th and $j$-th bases as $h$ and the
rest as identity, e.g., $h_{12}=h\otimes {\bf 1}_3\otimes...{\bf 1}_N$,
with ${\bf 1}_i$ the $n\times n$ identity matrix acting on the $i$-th basis.
The wave function in a given region, say $x_1<x_2<...<x_N$, is of the form
\be\label{Psi}
\ba{rcl}
\Psi&=&\displaystyle\sum_{\alpha_1,...,\alpha_N=1}^n
\phi_{\alpha_1,...,\alpha_N}(x_1,...,x_N)
e_{\alpha_1}\otimes...\otimes e_{\alpha_N}\\[3mm]
&=&u_{12...N}e^{i(k_1x_1+k_2x_2+...+k_Nx_N)}
+u_{21...N}e^{i(k_2x_1+k_1x_2+...+k_Nx_N)}+(N!-2)\, {\rm other~terms},
\ea
\ee
where $k_j$, $j=1,...,N$, are the momentum of the $j$-th particle.
$u$ are $n^N\times 1$ matrices.
The wave functions in the other regions are determined
from (\ref{Psi}) by the requirement of
symmetry (for bosons) or antisymmetry (for fermions).
Along any plane $x_i=x_{i+1}$, $i\in 1,2,...,N-1$, we have
\be\label{a1n}
u_{\alpha_1\alpha_2...\alpha_j\alpha_{j+1}...\alpha_N}=
Y_{\alpha_{j+1}\alpha_j}^{jj+1}
u_{\alpha_1\alpha_2...\alpha_{j+1}\alpha_j...\alpha_N},
\ee
where
\be\label{y}
Y_{\alpha_{j+1}\alpha_j}^{jj+1}=[2ik_{\alpha_j\alpha_{j+1}}-h_{j{j+1}}]^{-1}
[2jk_{\alpha_j\alpha_{j+1}}P^{jj+1} + h_{jj+1}].
\ee
Here $k_{\alpha_j\alpha_{j+1}}=(k_{\alpha_j}-k_{\alpha_{j+1}})/2$ play
the role of momenta
and $P^{jj+1}=p^{jj+1}$ for bosons and $P^{jj+1}=-p^{jj+1}$ for fermions,
where $p^{jj+1}$ is the operator on the $n^N\times 1$ column $u$
that interchanges the spins of particles $j$ and $j+1$.

For consistency $Y$ must satisfy the Yang-Baxter equation with
spectral parameter \cite{y,y2},
\be\label{ybe1}
Y^{m,m+1}_{ij}Y^{m+1,m+2}_{kj}Y^{m,m+1}_{ki}
=Y^{m+1,m+2}_{ki}Y^{m,m+1}_{kj}Y^{m+1,m+2}_{ij},
\ee
or
$$
Y^{mr}_{ij}Y^{rs}_{kj}Y^{mr}_{ki}
=Y^{rs}_{ki}Y^{mr}_{kj}Y^{rs}_{ij}
$$
if $m,r,s$ are all unequal, and
\be\label{ybe2}
Y^{mr}_{ij}Y^{mr}_{ji}=1,~~~~~~
Y^{mr}_{ij}Y^{sq}_{kl}=Y^{sq}_{kl}Y^{mr}_{ij}
\ee
if $m,r,s,q$ are all unequal.
By a straightforward  calculation it can be shown that the operator $Y$ given
by (\ref{y}) satisfies all the Yang-Baxter relations if
\be\label{hp}
[h_{ij}, P^{ij}]=0.
\ee
Therefore if the Hamiltonian operators for the spin coupling commute with the
spin permutation operator, the $N$-body quantum system (\ref{HN}) can be
exactly solved. The wave function is then given by (\ref{Psi}) and
(\ref{a1n}) with the energy $E=\displaystyle\sum_{i=1}^N k_i^2$.

For the case of spin-$1\over 2$, a Hermitian
matrix satisfying (\ref{hp}) is generally of the form
\be\label{hspin}
h^{1\over 2}=\left(\ba{cccc}a&e_1&e_1&c\\
e_1^\ast&f&g&e_2\\e_1^\ast&g&f&e_2\\
c^\ast&e_2^\ast&e_2^\ast&b\ea\right),
\ee
where $a,b,c,f,e_1,e_2\in\Cb$, $g\in \Rb$.
We recall that for a complex vector space  $V$, a matrix $R$
taking values in $End_c(V\otimes V)$ is called a solution of the
Yang-Baxter equation without spectral parameters, if it satisfies
\be\label{ybe}
{\cal R}_{12}{\cal R}_{13}{\cal R}_{23}=
{\cal R}_{23}{\cal R}_{13}{\cal R}_{12},
\ee
where ${\cal R}_{ij}$ denotes the matrix on the complex vector space
$V\otimes V\otimes V$, acting as $R$ on the $i$-th and the $j$-th
components and as identity on the other components. When $V$ is a two
dimensional complex space, the solutions of (\ref{ybe}) include
the ones such as $R_q$ which gives rise to the quantum algebra
$SU_q(2)$ and the integrable Heisenberg spin-$1\over 2$ chain models
such as the XXZ model ($R$ corresponds
to the spin coupling operator between the nearest neighbor spins in
Heisenberg spin chain models)\cite{ma2,ma3,ma1,ma4}.
Nevertheless in general $h^{1\over 2}$ does not satisfy the
Yang-Baxter equation without spectral parameters: $h^{1\over
2}_{12}h^{1\over 2}_{13}h^{1\over 2}_{23}\neq h^{1\over 2}_{23}h^{1\over
2}_{13}h^{1\over 2}_{12}$. But (\ref{hspin}) includes the Yang-Baxter
solutions, such as $R_q$, that
gives integrable spin chain models
(for an extensive investigation of the Yang-Baxter solutions see
\cite{ybe,jamo}). Therefore for an $N$-body system to be integrable,
the spin coupling in the contact interaction (\ref{HN}) is allowed to be
more general than the spin coupling in a Heisenberg spin chain model with
nearest neighbors interactions.

We now investigate the problem of bound states.
For $N=2$, from (\ref{a1}) the bound states have the form,
\be\label{bpsi2}
\psi^2_\alpha=u_\alpha e^{\frac{c+a\Lambda_\alpha}{2}\vert x_2-
x_1\vert},~~~~\alpha=1,...,n^2,
\ee
where $u_\alpha$ is the common $\alpha$-th eigenvector of $h$ and
$P^{12}$, with eigenvalue $\Lambda_\alpha$,
s.t. $hu_\alpha=\Lambda_\alpha u_\alpha$ and $c+a\Lambda_\alpha<0$,
$P^{12}u_\alpha=u_\alpha$.
The eigenvalue of the Hamitonian $H$ corresponding to the bound state
(\ref{bpsi2})
is $-(c+a\Lambda_\alpha)^2/2$. We remark that, whereas for the case of
the boundary condition (\ref{bound}), for a $\delta$ interaction one has
a unique bound state, here we have $n^2$ bound states.

By generalization we get the bound state for the $N$-particle system,
\be\label{bpsin}
\psi^N_\alpha=v_\alpha e^{-\frac{c+a\Lambda_\alpha}{2}\sum_{i>j}\vert
x_i-x_j\vert},~~~\alpha=1,...,n^2,
\ee
where $v_\alpha$ is the wave function of the spin part.

It can be checked that $\psi^N_\alpha$ satisfy the boundary condition
(\ref{b2}) at $x_i=x_j$ for any $i\neq j\in 1,...,N$. The spin wave
function $v$ here satisfies $P^{ij}v_\alpha=v_\alpha$ and
$h_{ij}v_\alpha=\Lambda_\alpha v_\alpha$, for any $i\neq j$.

It is worth mentioning that $\psi^N_\alpha$ is of the form (\ref{Psi})
in each of the above regions. For instance comparing $\psi^N_\alpha$
with (\ref{Psi}) in the region $x_1<x_2...<x_N$ we get
\be\label{k}
k_1=-i\frac{c+a\Lambda_\alpha}{2}(N-1),~k_2=k_1+ic,~k_3=k_2+ic,...,k_N=-k_1,
\ee
for $\alpha=1,...,n^2$. The energy of the bound state $\psi^N_\alpha$ is
\be\label{e}
E_\alpha=-\frac{(c+a\Lambda_\alpha)^2}{12}N(N^2-1).
\ee

Now we pass to the scattering matrix. For real $k_1<k_2<...k_N$, in
each coordinate region such as $x_1<x_2<...x_N$, the following term in
(\ref{Psi}) describes an outgoing wave
\be\label{out}
\psi_{out}=u_{12...N}e^{i(k_1x_1+...+k_Nx_N)}.
\ee
An incoming wave with the same exponential as (\ref{out}) is given by
\be\label{in}
\psi_{in}=[P^{1N}P^{2(N-1)}...]u_{N(N-1)...1}e^{i(k_Nx_N+...+k_1x_1)}
\ee
in the region $x_N<x_{N-1}<...<x_1$. From (\ref{a1n})
the scattering matrix $S$ defined by $\psi_{out}=S\psi_{in}$ is given by
\be\label{s}
S=[X_{21}X_{31}...X_{N1}][X_{32}X_{42}...X_{N2}]...[X_{N(N-1)}],
\ee
where $X_{ij}=Y^{ij}_{ij}P^{ij}$.

The scattering matrix $S$ is unitary and
symmetric due to the time reversal invariance of the interactions.
$<s_1^\p s_2^\p...s_N^\p\vert S\vert s_1s_2...s_N>$ stands for the $S$
matrix element of the process from the state
$(k_1s_1,k_2s_2,...,k_Ns_N)$ to the
state $(k_1s_1^\p,k_2s_2^\p,...,k_Ns_N^\p)$.

The scattering of clusters (bound states)
can be discussed in a similar way as in \cite{y1}.
For instance for the scattering of a bound state of two
particles ($x_1<x_2$) on a bound state of three particles
($x_3<x_4<x_5$), the scattering matrix is $S=[X_{32}X_{42}X_{52}]
[X_{31}X_{41}X_{51}]$.

The integrability of many particles system with contact spin coupling
interactions governed by separated boundary conditions can also be
studied. Instead of (\ref{BOUND}) we need to deal with the case
\be\label{bounds}
\phi^\prime(0_+) = G^+ \phi (0_+), ~~~
\phi^\prime(0_-) = G^- \phi (0_-) ,
\ee
where $G^{\pm}$ are Hermitian matrices.
For $G^+=G^-\equiv G$, $G^\dagger =G$, there is a Bethe Ansatz solution
to (\ref{Psi}) with $Y_{l_{i+1}l_i}^{ii+1}$ in (\ref{a1n}) given by
\be\label{ys}
Y_{l_{i+1}l_i}^{ii+1}=
\frac{ik_{l_il_{i+1}} + G}{ik_{l_il_{i+1}} - G}.
\ee

Let $\Gamma$ be the set of $n^2$ eigenvalues of $G$. For any
$\lambda_\alpha\in \Gamma$ such that $\lambda_\alpha<0$, there are
$2^{N(N-1)/2}$ bound states for the $N$-particle system,
\be\label{bpsins}
\psi^{N}_{\alpha\underline{\epsilon}}=
v_{\alpha\underline{\epsilon}}
\prod_{k>l} (\theta (x_k-x_l) +\epsilon_{kl}\theta (x_l-x_k))
e^{\lambda_\alpha\sum_{i>j} \vert x_i-x_j\vert },
\ee
where $v_{\alpha\underline{\epsilon}}$ is the spin wave function and
$\underline{\epsilon} \equiv \{ \epsilon_{kl}~:~k>l \}$; $\epsilon_{kl}=\pm$,
labels the $2^{N(N-1)/2}$-fold degeneracy.
The spin wave
function $v$ here satisfies $P^{ij}v_{\alpha\underline{\epsilon}}
=\epsilon_{ij}v_{\alpha\underline{\epsilon}}$
for any $i\neq j$, that is, $p^{ij}v_{\alpha\underline{\epsilon}}
=\epsilon_{ij}v_{\alpha\underline{\epsilon}}$
for bosons and $p^{ij}v_{\alpha\underline{\epsilon}}
=-\epsilon_{ij}v_{\alpha\underline{\epsilon}}$ for fermions.

Again $\psi^{N}_{\alpha\underline{\epsilon}}$ is of the form (\ref{Psi})
in each of the regions $x_{i_1}<x_{i_2}<...<x_{i_N}$.
For instance comparing $\psi^{N}_{\alpha\underline{\epsilon}}$ with
(\ref{Psi}) in the region $x_1<x_2...<x_N$ we get
$k_1=i\lambda_\alpha (N-1)$, $k_2=k_1-2i\lambda_\alpha$,
$k_3=k_2-2i\lambda_\alpha$,...,$k_N=-k_1$.
The energy of the bound state $\psi^{N}_{\alpha\underline{\epsilon}}$ is
\be\label{es}
E_\alpha=-\frac{\lambda_\alpha^2}{3}N(N^2-1).
\ee

We have investigated the integrable models of $N$-body systems with
contact spin coupling interactions. Without taking into account the spin
coupling, the boundary condition (\ref{bound}) is characterized by four
parameters (separated boundary conditions are a special limiting
case of these). Obviously the general boundary condition (\ref{BOUND}) we
considered in this article has much more parameters. The classification
of the dynamic operators associated with different parameter regions
is a big challenge. As we have seen, the case $A=D={\bf I}_2$, $B=0$, $C=h$
corresponds to a Hamiltonian with $\delta$-interactions of the form
(\ref{H}) (for $N=2$). It can
be further shown that (for $N=2$) the following boundary condition
\be\label{BOUND1}
\left( \begin{array}{c}
\psi\\
\psi '\end{array} \right)_{0^+}
=\left(
\begin{array}{cc}
{\bf I} & B \\
0 & {\bf I} \end{array} \right)
\left( \begin{array}{c}
\psi\\
\psi '\end{array} \right)_{0^-},
\end{equation}
corresponds to a Hamiltonian $H$ of the form:
$$
H=-D^2_x(1+B\delta) - BD_x\delta^\prime,
$$
where $B$ is an $n^2\times n^2$ Hermitian matrix,
$D_x$ is defined by $(D_x f)(\varphi)=-f(\frac{d}{dx}\varphi)$, for
$f\in C^\infty_0(\Rb\slash\{0\})$ and $\varphi$ a test function with a
possible discontinuity at the origin.

The boundary condition
\be\label{BOUND2}
\left( \begin{array}{c}
\psi\\
\psi '\end{array} \right)_{0^+}
=\left(
\begin{array}{cc}
\frac{2+iB}{2-iB} & 0 \\
0 & \frac{2-iB}{2+iB} \end{array} \right)
\left( \begin{array}{c}
\psi\\
\psi '\end{array} \right)_{0^-},
\end{equation}
describes the Hamiltonian
$$
H=-D_x^2+iB(2D_x\delta-\delta^\prime).
$$

We have introduced boundary conditions depending on the spin states of
the particles and studied several special cases. A complete investigation
of integrable $N$-body systems and Hamiltonian operators corresponding
to the general boundary conditions of the form (\ref{BOUND})
still remains to be done.


\begin{thebibliography}{99}

\bibitem{abd}
S. Albeverio, Z. Brze\'{z}niak and L. D\c{a}browski,
Time-dependent propagator with point interaction,
{\it J. Phys.} {\bf  A 27} (1994)4933-4943.

\bibitem{15b}
S. Albeverio, Z. Brze\'{z}niak and L. D\c{a}browski,
Fundamental somution of heat and Schr\"odinger
equations with point interaction, {\it J.Funct.Anal.},
{\bf 130} (1995)220-254.

\bibitem{ADF}
S. Albeverio, L. D{\c a}browski and S.M. Fei,
          {\it  One Dimensional Many-Body Problems with Point
          Interactions}, quant-ph/0001089, to appear in {\sf Int.
          J. Mod. Phys. B}.

\bibitem{adk}
S. Albeverio, L. D\c{a}browski and P. Kurasov,
Symmetries of Schr\"odinger operators with point
interactions,
{\it Lett. Math. Phys.}, {\bf 45} (1998)33-47.

\bibitem{AFK}
S. Albeverio, S.M. Fei and P. Kurasov,
          {\it Gauge Fields, Point Interactions and Few-Body Problems
          in One Dimension}, preprint 1999.

\bibitem{agh-kh}
S. Albeverio, F. Gesztesy, R. H\o egh-Krohn and H. Holden, {\it
Solvable Models in Quantum Mechanics}, New York: Springer, 1988.

\bibitem{albku}
S.Albeverio and P.Kurasov,
Finite rank perturbations and distribution
theory,
{\it Proc. AMS}, {\bf 127} (1999)1151-1161.

\bibitem{AKbook}
S. Albeverio and R. Kurasov, {\it Singular perturbations of differential
operators and solvable Schr\"odinger type operators},
Cambridge Univ. Press., to appear in 1999.

\bibitem{buslaev}
V. S. Buslaev, S. P. Merkuriev, and S. P. Salikov,
{\it On the diffractional character of the scattering problem for three
one-dimensional particles},
Problemu Mat.  Fiz., vup.9, Leningrad University Press (1979) (in
Russian).

\bibitem{ma2}
V. Chari and A. Pressley, {\it A Guide to Quantum Groups}, Cambridge
University Press, 1994.

\bibitem{ch} P.R. Chernoff and R.J. Hughes,
A new class of point interactions in one dimension,
 {\it J. Funct. Anal.}, {\bf 111 }(1993)97-117.

\bibitem{coutinho} F.A.B.Coutinho, Y.Nogami, and Lauro Tomio,
Many--body system with a four--parameter family of point
interactions in one dimension,
{\it J.Phys.A}, {\bf 32} (1999)4931-4942.

\bibitem{ybe}
S.M. Fei, H.Y. Guo and H. Shi, {\it Multiparameter
	  Solutions of the Yang-Baxter Equation},
	  {\it J. Phys.} {\bf A 25}(1992)2711-2720.

\bibitem{gaudin}
M. Gaudin, {\it La fonction d'onde de Bethe}, Masson, 1983.

\bibitem{y1}
C.H. Gu and C.N. Yang, {\it A one-dimensional $N$ Fermion
problem with factorized $S$ matrix}, Commun. Math. Phys. {\bf 122}
(1989)105-116.

\bibitem{jamo}
J. Hietarinta, {\it All solutions to the constant quantum Yang-Baxter
equation in two dimensions}, {\it Phys. Lett.} A {\bf 165}(1992),
245-251.

\bibitem{ma3}
C. Kassel, {\it Quantum Groups}, Springer-Verlag, New-York, 1995.

\bibitem{makarov}
Yu.A Kuperin, K.A. Makarov and B.S. Pavlov,
One-dimensional model of three-particle resonances,
{\it Teoret. Mat. Fiz.}, {\bf  63} (1985), 78-87.

\bibitem{kurasov}
P. Kurasov,
Distribution theory with disconstinuous test functions
and differential operators with generalized coefficients,
{\it J. Math. Analys. Appl.}, {\bf 201} (1996), 297-333.

\bibitem{revmath} P.Kurasov,
{ Energy dependent boundary conditions and few-body scattering
problem}, {\it Rev. in Math. Physics}, {\bf 9} (1997), 853-906.

\bibitem{jabo}
P.Kurasov and J.Boman, Finite rank singular perturbations and
distributions with discontinuous test functions,
{\it Proc. AMS}, {\bf 126} (1998), 1673--1683.

\bibitem{ma1}
Z.Q. Ma, {\it Yang-Baxter Equation and Quantum Enveloping Algebras},
World Scientific, 1993.

\bibitem{ma4}
S. Majid, {\it Foundations of Quantum Group Theory}, Cambridge
University Press, 1995.

\bibitem{mcguire} J.B.McGuire, Study of exactly soluble
one--dimensional N--body problems,
{\it J.Math.Phys.}, {\bf 5} (1964), 622--636.

\bibitem{mcguire2} J.B.McGuire, Interacting fermions in one
dimension.I. Repulsive potential,
{\it J.Math.Phys.}, {\bf 6} (1965), 432--439.

\bibitem{mcguire3} J.B.McGuire, Interacting fermions in one
dimension.II. Attractive potential,
{\it J.Math.Phys.}, {\bf 7} (1966), 123--132.

\bibitem{mcguire4} J.B.McGuire and C.A.Hurst,
The scattering of three impenetrable particles in one dimension,
{\it J.Math.Phys.}, {\bf 13} (1972), 1595--1607.

\bibitem{mcguire5} J.B.McGuire and C.A.Hurst,
Three interacting particles in one dimension:
an algebraic approach,
{\it J.Math.Phys.}, {\bf 29} (1988), 155--168.

\bibitem{seba} P. \v{S}eba, The generalized point interaction in one
dimension,
{\it Czechoslovak J. Phys. B}, {\bf 36 } (1986), 667--673.

\bibitem{y}
C.N. Yang, {\it Some exact results for the many-body problem
in one dimension with repulsive delta-function interaction},
{\it Phys. Rev.
Lett.}, {\bf 19} (1967), 1312-1315.

\bibitem{y2}
C.N. Yang, {\it $S$ matrix for the one-dimensional $N$-body problem with
repulsive $\delta $-function interaction},
{\it Phys. Rev. },
{\bf 168}(1968)1920-1923.

\end{thebibliography}
\end{document}